\documentclass[prl,twocolumn,superscriptaddress,showpacs,floatfix,longbibliography]{revtex4-1}
\usepackage{mathrsfs}
\usepackage{amssymb, amsbsy, amsmath, latexsym, dsfont, array, layout,mathrsfs,color,bm}
\usepackage{xcolor}
\usepackage{cancel}

\newcommand{\ket}[1]{\left|{#1}\right\rangle}

\usepackage{graphicx}
\definecolor{delete}{rgb}{1.0, 0.0, 0.0}
\definecolor{edit}{rgb}{0.0, 0.0, 0.9}
\definecolor{comment}{rgb}{0.9, 0.0, 0.0}

\newcommand{\mh}[1]{{\textcolor{comment}{#1}}}

\begin{document}

\title{Non-Hermitian Kibble-Zurek mechanism with  tunable complexity \\ in single-photon interferometry}
\author{Peng Xue}\email{gnep.eux@gmail.com}
\author{Lei Xiao}
\affiliation{Beijing Computational Science Research Center, Beijing 100084, China}
\author{Dengke Qu}
\affiliation{Department of Physics, Southeast University, Nanjing 211189, China}
\affiliation{Beijing Computational Science Research Center, Beijing 100084, China}
\author{Kunkun Wang}
\affiliation{Beijing Computational Science Research Center, Beijing 100084, China}
\author{Hao-Wei Li}
\author{Jin-Yu Dai}
\affiliation{CAS Key Laboratory of Quantum Information, University of Science and Technology of China, Hefei 230026, China}
\author{Bal\'azs D\'ora}
\affiliation{Department of Theoretical Physics and MTA-BME Lend\"ulet Topology and Correlation Research Group,
Budapest University of Technology and Economics, 1521 Budapest, Hungary}
\author{Markus Heyl}
\affiliation{Max-Planck-Institut f\"ur Physik komplexer Systeme, 01187 Dresden, Germany}
\author{Roderich Moessner}
\affiliation{Max-Planck-Institut f\"ur Physik komplexer Systeme, 01187 Dresden, Germany}
\author{Wei Yi}
\affiliation{CAS Key Laboratory of Quantum Information, University of Science and Technology of China, Hefei 230026, China}
\affiliation{CAS Center For Excellence in Quantum Information and Quantum Physics,Hefei 230026, China}

\begin{abstract}
{\bf
Non-Hermitian descriptions of quantum matter have seen impressive progress recently~\cite{ptreview,zeuner,XZB+17,ZXB+17,luole,gadway,bender13,rivet,yoshida,stehmann,SLZ+11,FSA15,E19}, with major advances in understanding central aspects such as their topological properties or the physics of exceptional points, the non-Hermitian counterpart of critical points~\cite{heiss}.
Here, we use single-photon interferometry to reconstruct the non-Hermitian Kibble-Zurek mechanism and its distinct scaling behavior for exceptional points~\cite{dhm}, by simulating the defect production upon performing slow parameter ramps.
Importantly, we are able to realise also higher-order exceptional points, providing experimental access to their theoretically predicted  characteristic Kibble-Zurek scaling behaviour.
Our work represents a crucial step in increasing the experimental complexity of non-Hermitian quantum time-evolution. It thus also furthers the quest to move the frontier from purely single-particle physics towards increasingly complex settings in the many-body realm.
}
\end{abstract}
\date{\today}

\maketitle

The foundational axioms of quantum mechanics impose a Hermitian structure on Hamiltonians.
However, it is now appreciated that rich and unconventional phenomena can arise in settings where
the constraints enforced by such Hermitian structures are absent. This happens rather generically for
systems in touch with an environment; experimental instances occurring in
photonics~\cite{ptreview,zeuner,XZB+17,ZXB+17}, cold atoms~\cite{luole,gadway}, mechanical systems~\cite{bender13,rivet,yoshida}, and electric circuits~\cite{stehmann,SLZ+11,FSA15,E19}
have revealed rich single-particle properties induced by non-Hermiticity.
In particular, the so-called exceptional points (EPs)~\cite{heiss,dhm,HHW+17,ORNY19} provide the
non-Hermitian~\cite{bender98}
counterpart to the familiar critical points~\cite{sachdev}
in Hermitian band structures~\cite{KAU+17,xiao}, leading to critical phenomena unique to non-Hermitian systems.
Whereas most previous studies focus on single-particle properties,
we are now faced with the challenge of extending their experimental reach, by realising specific phenomena characteristic of non-unitary dynamics on one hand, and allowing the study of more complex problems on the other, shifting the frontier towards the many-body realm.

In this work, we report a progress along both axes using single-photon interferometric networks as an experimental platform.
We provide a framework to achieve highly tunable non-Hermitian band structures allowing us to realise different classes of EPs with varying complexity. Mathematically, at an EP, two (or more) complex eigenvalues and eigenstates coalesce~\cite{heiss}. These eigenstates then no longer form a complete basis, which in turn feeds into distinct properties of the EP compared to those at Hermitian critical points.
We consider various types of non-Hermitian EPs, culminating in a higher-order EP, the counterpart of the familiar higher-order critical points, which generically appear in multi-dimensional phase diagrams upon tuning a pair of parameters.
%
%

Such a framework enables us to systematically characterise the novel non-unitary dynamics pertaining to different types of EPs, which is specifically visible in the dynamics of defect generation upon passage through the exceptional/critical point in the time domain, as captured by the venerable Kibble-Zurek mechanism~\cite{kibble,zurek,XHS+14,ARB+16,LK16,KOL+19}.
We simulate such ramps in a many-body setting, and probe the characteristic universal scaling behavior of the resulting defect density:
\begin{equation}
        n \sim v^{-d^\ast\nu/(z\nu +1)} \, .
        \label{eq:KZscaling}
\end{equation}
Here, $v$ denotes the sweep velocity and $\nu$ the correlation length exponent.
Remarkably, the defect density $n$ obeys a scaling form even in the non-Hermitian case with, however, one crucial difference.
While for conventional critical points $d^\ast = d$ is identical to the spatial dimension $d$, in the non-Hermitian setting the effective dimension~\cite{dhm} $d^\ast = d + z$ contains a shift by the dynamical critical exponent $z$.
%
Using single-photon interferometric networks we measure $n$ and observe a power-law dependence with exponents in keeping with the Kibble-Zurek prediction both for the Hermitian and the non-Hermitian cases.

With the ability to tune the band structure complexity, we then also address the Kibble-Zurek mechanism for higher-order EPs and determine their characteristic properties experimentally.
Our approach is based on simulating sets of independent modes, mimicking the dynamics of effectively noninteracting quasiparticles upon crossing critical and exceptional points.
This technique provides a flexible framework for realising and probing non-unitary dynamics with high tunability and control, with the potential to push the frontier further towards the quantum simulation of many-body effects.

\begin{figure*}[tbp]
\includegraphics[width=0.8\textwidth]{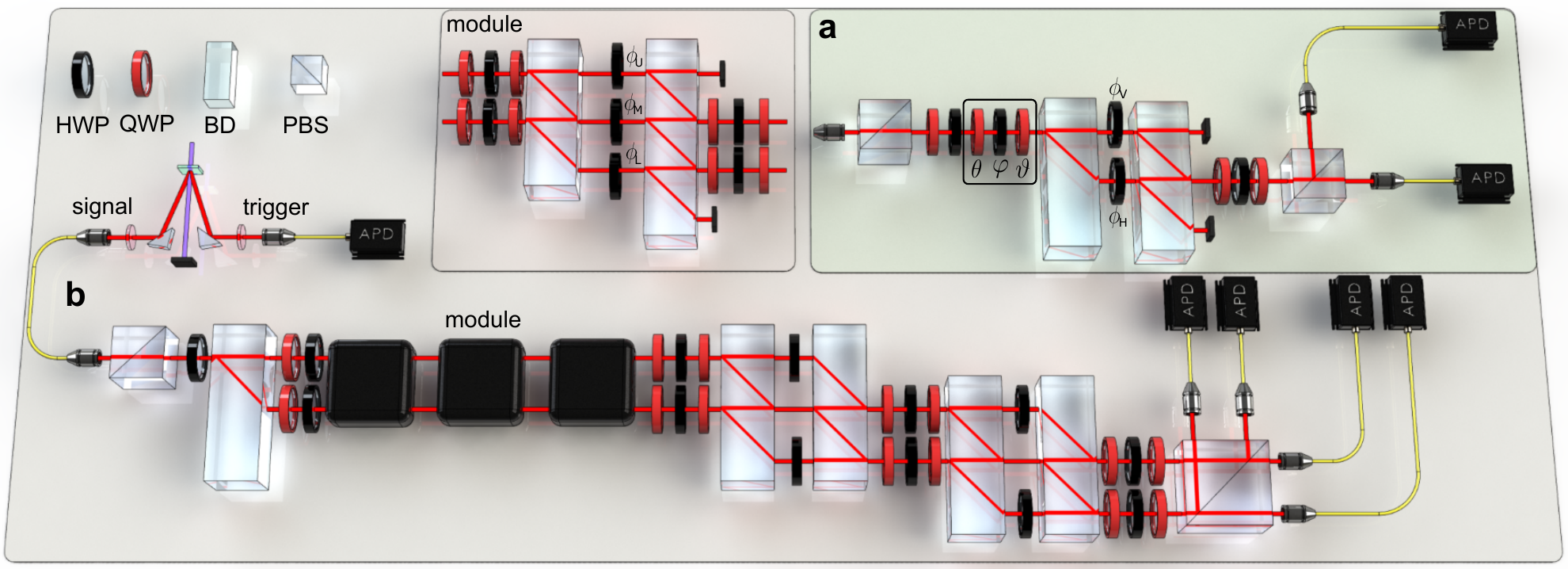}
\caption{{\bf Experimental setup.} As a first step a photon pair is created via spontaneous parametric down-conversion (SPDC). One of the photons serves as a trigger, the other is projected into the polarisation state $\ket{H}$ or $\ket{V}$ with a polarising beam splitter (PBS) and a half-wave plate (HWP) before entering the interferometric network. {\bf a} Experimental setup for the non-unitary dynamics of a single-qubit $\mathcal{PT}$-symmetric system. {\bf b} Experimental setup for the non-unitary dynamics of a four-level system, where the qudit is encoded in polarisational and spatial degrees of freedom. To prepare the initial state, heralded single photons pass through a PBS and a HWP with tailored setting angles and are split by a birefringent calcite beam displacer (BD) into two parallel spatial modes: upper and lower modes. After passing through wave plates inserted into optical paths of the two spatial modes, the photons are prepared in the initial state $\ket{\Psi_p(0)}$ or $\ket{\varphi_p}$ (see the main text for definitions), through $p$-dependent wave plates. The non-unitary evolution is then realised by three modules involving BDs and wave plates, and projective measurements are performed by a cascaded interferometer and avalanche photodiodes (APDs).
}
\label{fig:setup}
\end{figure*}

~\\
\noindent
{\bf Results}
~\\
\noindent
{\bf Model, observables and protocol}
\\
\noindent
We consider translationally invariant quantum systems which can be described in terms of independent modes:
\begin{align}
H=\sum_p H_p,\quad H_p=p\sigma_x+\Delta\sigma_y+i\Gamma\sigma_z \, .
\label{ham2x2}
\end{align}
Here, $H_p$ denotes a $2\times2$ matrix represented in terms of Pauli operators $\sigma_\alpha$ ($\alpha=x,y,z$), and parametrized through the momentum $p$ of the mode and the couplings $\Delta,\Gamma$.
This Hamiltonian exhibits genuine non-Hermitian character due to the complex mass term, with $\Gamma \in \mathbb{R}$ implying $H^\dag \not= H$.
The resulting spectrum $E_\pm(p)=\pm\sqrt{p^2+\Delta^2-\Gamma^2}$ is gapless at $p^2+\Delta^2=\Gamma^2$, which marks the location of a second-order EP.
It is the key goal of this work to experimentally study the dynamical consequences of such EPs~\cite{doppler} and to contrast with those of conventional Hermitian critical points~\cite{sachdev,KOL+19}.

Based on the general protocol that we outline below, the non-unitary dynamics can be further enriched by engineering $H_p$ as an enlarged $4\times 4$ matrix, enabling us to access a mode structure of increased complexity, and the concomitant unconventional higher-order EPs.

For our protocol, we initialise each mode $p$ in its ground state $|\Psi_p\rangle$ with $\Gamma=0$.
At time $t=0$, we then start the non-equilibrium process by linearly increasing $\Gamma \propto t$, driving the system either through a critical or exceptional point.
The defects in the final state $|\Psi_p(\tau)\rangle$ at time $t=\tau$ are quantified via
\begin{equation}
\sigma_z(p,\tau)=\frac{\langle\Psi_p(\tau)|\sigma_z|\Psi_p(\tau)\rangle}{\langle\Psi_p(\tau)|\Psi_p(\tau)\rangle} \,.
\end{equation}
The total defect density
$n=(2\pi)^{-1}\int \text{d}p \sigma_z(p,\tau)$ according to the Kibble-Zurek prediction Eq.~(\ref{eq:KZscaling}) shows universal behavior,
when measured with respect to its equilibrium or steady-state value, $n_{\rm eq}$ (see Methods). For ease of
notation, we set the velocity of $p$-modes and $\hbar$ to unity, and the  density of states to $1/\pi$.

\begin{figure*}[tbp]
\includegraphics[width=0.8\textwidth]{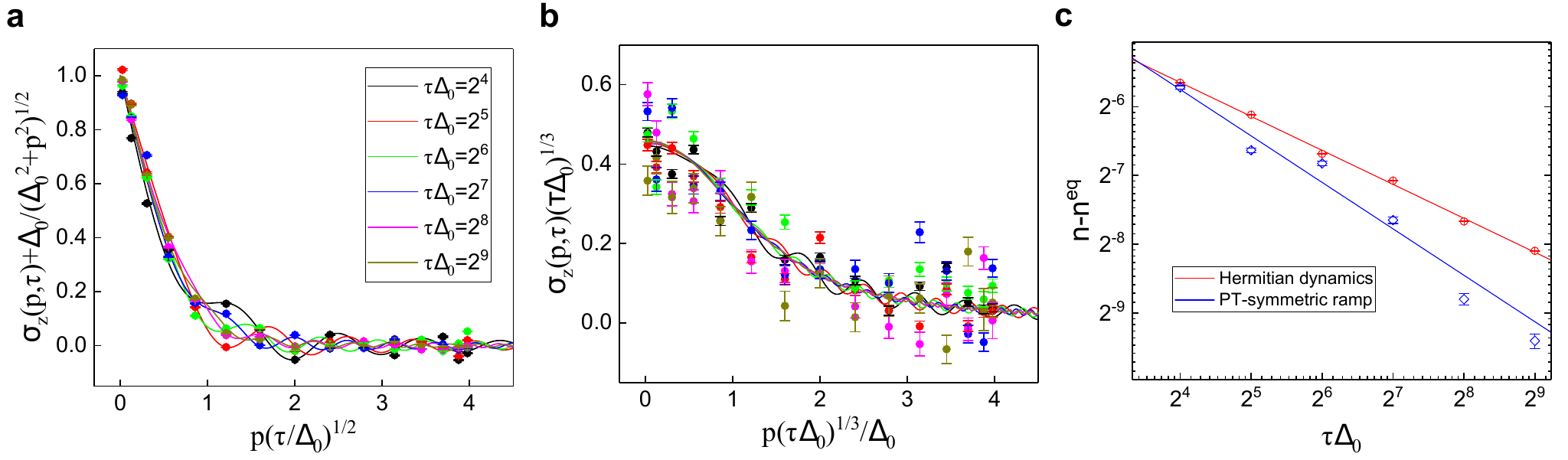}
\caption{{\bf Kibble-Zurek scaling of defect production for parametric ramps in Hermitian and $\mathcal{PT}$-symmetric non-Hermitian systems.} Data collapse for the momentum-resolved defect density for the Hermitian case {\bf a} and the $\mathcal{PT}$-symmetric ramp {\bf b}. Dots with error bars display the experimental measurements and solid lines refer to numerical simulations. {\bf c} Total defect density $n$ relative to the adiabatic values $n_{\rm eq}$ for the two considered cases as a function of the ramp time $\tau$. Solid lines show the results for the best power-law fit $n-n_{\rm eq}\sim \tau^{-\alpha}$ to the experimental data. The fitted exponents $\alpha=0.494(2)$ for the Hermitian scaling (red), and $\alpha=0.68(1)$ for the $\mathcal{PT}$-symmetric ramp (blue).
}
\label{fig:fig2}
\end{figure*}

\begin{figure}[tbp]
\includegraphics[width=0.5\textwidth]{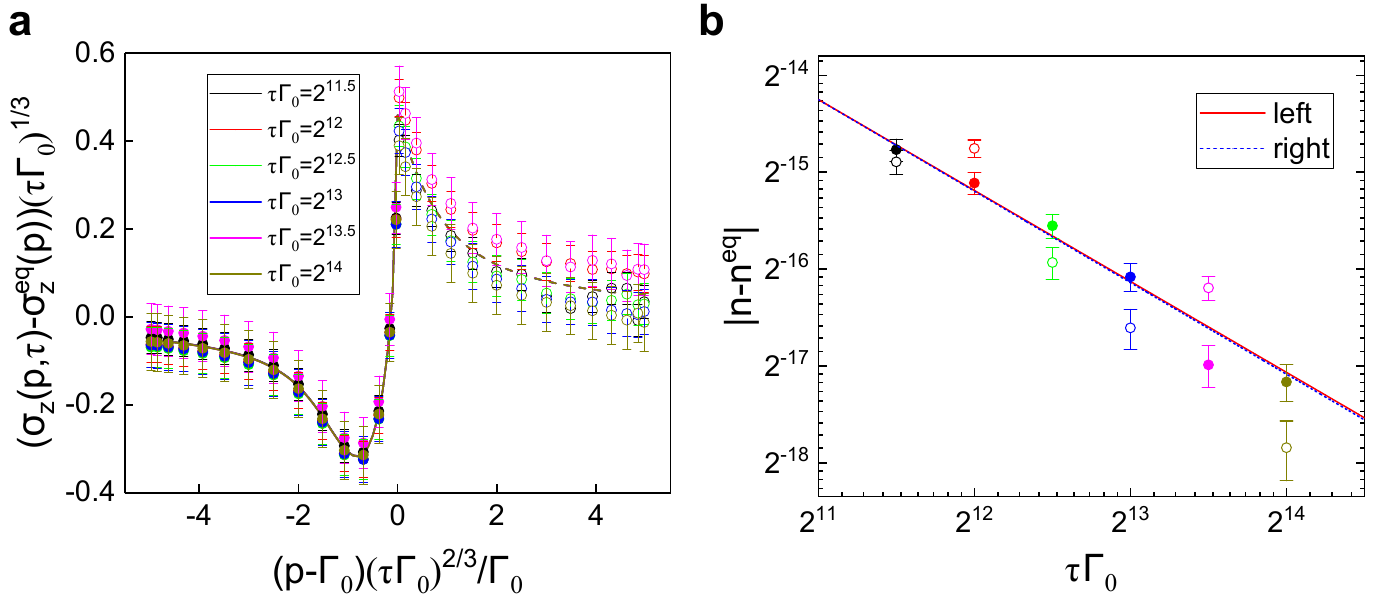}
\caption{{\bf Kibble-Zurek scaling of defect production for parameter ramps across non-Hermitian EPs.} Data collapse for the momentum-resolved defect density for the full non-Hermitian drive {\bf a}.
Total defect density relative to the adiabatic values as a function of the ramp time $\tau$ {\bf b}. As discussed in the main text, we sample the left ($p<\Gamma_0$) and right ($p>\Gamma_0$) regions separately. The red solid (black dashed) line in {\bf b} shows the results for the best power-law fits $|n-n_{\rm eq}|\sim \tau^{-\alpha}$ for the left (right) region, with the fitted exponents $\alpha=0.94(7)$ in both regions.
}
\label{fig:fig3}
\end{figure}

~\\
\noindent
{\bf Experimental implementation}
\\
\noindent
Experimentally, we realise for each mode $p$
the time-evolution operator of the targeted $H_p$
in single-photon interferometry. The basis states are encoded in the horizontal and vertical
polarisations of single photons, with $\ket{H}=(1,0)^\text{T}$ and $\ket{V}=(0,1)^\text{T}$. The photons are initialised in the
ground state $\ket{\Psi_p(0)}$ of $H_p$ with $\Gamma=0$ through a series of wave plates with $p$-dependent parameters (see Methods),
and are sent to the interferometric network as illustrated in Fig.~\ref{fig:setup}.
The associated time-evolution operator $U_p(\tau)$ is realised by  three successive operations
$U_p(\tau)=R(\theta_2,\varphi_2,\vartheta_2)L(\theta_H,\theta_V)R(\theta_1,\varphi_1,\vartheta_1)$.
Here, the rotation operators $R(\theta_j,\varphi_j,\vartheta_j)$ ($j=1,2$) are experimentally implemented using a set of sandwich-type wave plates with setting angles $\varphi_j$, $\theta_j$, and $\vartheta_j$. The polarisation-dependent loss operator
$L(\theta_H,\theta_V$) contributes the non-unitary dynamics, and is experimentally achieved by a combination of beam displacers (BDs) and HWPs with setting angles $\phi_H$
and $\phi_V$. We choose the parameters $\{\theta_j,\varphi_j,\vartheta_j,\phi_H,\phi_V\}$ such the sequence of three operations reproduces the targeted $U_p(\tau)$. Such a scheme enables us to implement any desired non-unitary operator for
two-level systems~\cite{xiao}.
We measure $\sigma_z(p,\tau)$ by recording the relative photon counts in the basis $\{\ket{H},\ket{V}\}$ through a PBS and avalanche photodiodes (APDs), with a typical peak count of $160,000$ photons.
In order to estimate the total defect density $n$, we utilize Gauss-Legendre quadrature which uniquely determines the specific $p$ points once the integration domain is fixed.

~\\
\noindent
{\bf Hermitian Kibble-Zurek scaling}
\\
\noindent
We first validate the set-up by reproducing the well-known Kibble-Zurek scaling for Hermitian quantum critical points with quantitative accuracy.
Concretely, we consider  $\Delta=0$ and $\Gamma=-i\Delta_0 t/\tau$, yielding a mass term linearly increasing with time.
Figure \ref{fig:fig2}a shows the results for $\sigma_z(p,\tau)$ in scaled units so as to achieve a data collapse with the predicted exponents:
the resulting total defect density $n$, see Fig.~\ref{fig:fig2}c, is consistent with power-law behavior over more than one decade.
From a fit $n-n_{\rm eq}\sim \tau^{-\alpha}$ we obtain $\alpha=0.494(2)$, which, within the error
bars, agrees with the theoretical prediction $\alpha=0.5$ for the underlying equilibrium quantum critical point of Ising universality class with $d=z=1$, $\nu=1/2$.
Here, the constant $n_{\rm eq}$ denotes the expected adiabatic value for $\sigma_z(p,\tau)$ in the $\tau\rightarrow\infty$ limit.

~\\
\noindent
{\bf Non-Hermitian Kibble-Zurek scaling}
\\
\noindent
For the central aspect of this work, the non-Hermitian Kibble-Zurek scaling at EPs, we consider two cases.
First, a parity-time ($\mathcal{PT}$)-symmetric ramp with $\Delta=\Delta_0$ and $\Gamma=\Delta_0 t/\tau$, see Figs.~\ref{fig:fig2}b and c, where the energy spectrum $E_\pm(p)$ of the system remains real throughout the ramp approaching an EP at $t=\tau$.
Second, a fully non-Hermitian drive with $\Delta=0$, $\Gamma=\Gamma_0t/\tau$, see Fig.~\ref{fig:fig3}.
To reduce experimental error for the fully non-Hermitian drive, we sample the mode-resolved defect densities and analyse the scaling behaviour in
regions $p\gtrless\Gamma_0$  separately. Whereas the difference in the integrated densities between the two
regions gives the total integrated density as before, the smallness of such a difference would lead to small
photon counts and significantly larger error bars. By analysing the two regions separately, as we show below, a
universal scaling behaviour is established, since both regions respect the same scaling law.

%
%
For both cases, the data collapse across the varying ramp times $\tau$, characteristic of the Kibble-Zurek mechanism, but now realised for non-Hermitian systems.
The resulting scaling functions, however, differ markedly from the Hermitian case in Fig.~\ref{fig:fig2}a.
Especially, the full non-Hermitian drive provides a manifestly non-Hermitian feature:
the double-peak structure in Fig.~\ref{fig:fig3}a implies modes $p$ with lower occupation than for the purely adiabatic limit, which is impossible for the Hermitian case.
%

%

%
While the scaling functions appear to have rather unconventional form, the integrated total defect density $n$ robustly exhibits power-law behaviour, as illustrated in Figs.~\ref{fig:fig2}c and \ref{fig:fig3}b.
The associated exponents follow the modified scaling law in Eq.~(\ref{eq:KZscaling}) with the effective dimension $d^\ast$.
For the $\mathcal{PT}$-symmetric ramp, the fit $n-n_{\rm eq}\sim \tau^{-\alpha}$ yields an exponent $\alpha=0.68(1)$, consistent with the theoretical prediction $\alpha=2/3$ for the underlying EP with critical exponents $d=z=1$, $\nu=1/2$.
For the full non-Hermitian drive, the fitted exponents are both $\alpha=0.94(7)$ for the two regions with $p\gtrless\Gamma_0$, see
Fig.~\ref{fig:fig3}b. These suggest that the total defect density should obey a power-law scaling in agreement with Eq.~(\ref{eq:KZscaling}) for $d=\nu=1$ and $z=1/2$.
These results demonstrate the high accuracy with which our experiments can probe the dynamics of non-Hermitian systems.

The increasing experimental error bars for longer evolution times in the non-unitary dynamics, Figs.~\ref{fig:fig2}c and \ref{fig:fig3}b, are mainly due to the non-unitary photon losses:
a smaller number of photons implies a larger statistical error.
Importantly, the influence of imperfections appears weaker for the total defect density $n$ than for the mode-resolved one $\sigma_z(p,\tau)$.
This happens because the errors in different modes are statistically independent, and hence suppressed upon integration:
the majority of experimental imperfections originate from wave plates and BDs that are independently tuned in different $p$ sectors.

~\\
\noindent
{\bf Kibble-Zurek at higher-order exceptional points}
\\
\noindent
The key next step is to increase the complexity of non-Hermitian Hamiltonians, accessing previously unreachable physical properties.
Here, we achieve this by enlarging the mode matrix $H_p$ to be of $4\times 4$ form, providing an experimental access to
a higher-order EP. Here,
\begin{align}
H^{(4)}_p=\begin{pmatrix}
                     0 & \Delta-\Gamma & 0 & ip \\
                     \Delta+\Gamma & 0 & \Delta-\Gamma & 0 \\
                     0 & \Delta+\Gamma & 0 & \Delta-\Gamma \\
                     -ip & 0 & \Delta+\Gamma & 0
                   \end{pmatrix}
\label{ep4ham}
\end{align}
For $\Gamma=0$, the model is Hermitian, while for $\Gamma=\Delta$, it is $\mathcal{PT}$-symmetric and features a fourth-order EP at $p = 0$.
Here the spectrum scales as $\sim p^{1/4}$ due to detuning from the EP, while the gap scales $\sim \sqrt{\Gamma-\Delta}$ at $p=0$, suggesting
critical exponents $z=1/4$ and $z\nu=1/2$.


The four basis states are now encoded in the polarisations and spatial modes of single photons,
and given
by $\{|UH\rangle=(1,0,0,0)^\text{T},|UV\rangle=(0,1,0,0)^\text{T},|DH\rangle=(0,0,1,0)^\text{T},|DV\rangle=(0,0,0,1)^\text{T}\}$. Here $|U\rangle$ and $|D\rangle$
represent, respectively, the upper and lower spatial modes of photons (see Fig.~\ref{fig:setup}b).
The experimental implementation of the calculated $U^{(4)}_p$, however, is different from that of $U_p$ for two-level systems which is realised {\it exactly} by BDs
and wave plates. Here, instead,
we approximate $U^{(4)}_p$ with a series of modules, each consisting of two BDs and a set of wave plates (see Methods for details).

\begin{figure}
\includegraphics[width=0.5\textwidth]{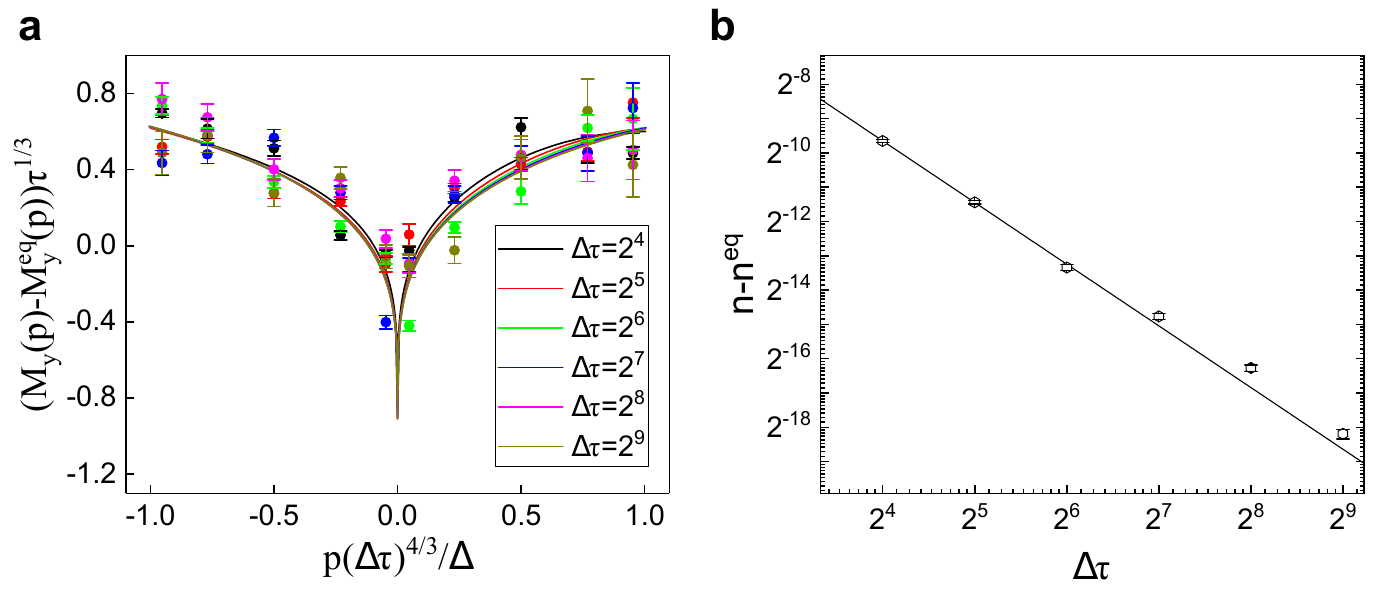}
\caption{{\bf Kibble-Zurek scaling for a higher-order EP.} Data collapse for the  momentum-resolved defect density for different ramp times $\tau$ {\bf a}. Dots with error bars refer to the experimental measurements and solid lines depict the exact numerical solution. {\bf b} The total defect density as a function of the ramp time $\tau$ exhibits algebraic behavior {\bf b}. The solid line represents the best power-law fit to the experimental data, with the fitted exponent $\alpha=1.80(5)$.
By moving further away from EP4, deviations from the ideal scaling for larger values of $p$ would appear.
This can be attributed to the additional EP2's at infinity, which imply levels being very close to each
other pairwise for any finite $p$, leading to inter-level tunnelings.
}
\label{fig:fig4}
\end{figure}

We characterize the final momentum-resolved defect density through $M_y(p)-M^{\text{eq}}_y(p)$, where
$M_y(p)=\langle \Psi_p(\tau)|M_y|\Psi_p(\tau)\rangle/\langle \Psi_p(\tau)|\Psi_p(\tau)\rangle$,
$M_y^\text{eq}(p)=\langle \varphi_p|M_y| \varphi_p\rangle \approx -\sqrt{1+\sqrt 2}|p|^{1/4}$ for small $p$ to a very good approximation, with $M_y$ given by $M_y=i\frac{\partial H^{(4)}_p}{\partial \Gamma}$ and
$|\varphi_p\rangle$ the eigenstate of $H^{(4)}_p$ with the largest imaginary eigenvalue under the condition $\Gamma=\Delta$.

The measured $M_y(p)-M^{\text{eq}}_y(p)$ for various $\tau$, again exhibits a scaling collapse, see Fig.~\ref{fig:fig4}a.
The defect production in the presence of a higher-order critical point exhibits a distinct feature, different from the previously studied cases in Figs.~\ref{fig:fig2}b and \ref{fig:fig3}a.
Specifically, $M_y(p)-M^{\text{eq}}_y(p)$ develops a pseudogap-like feature around $p=0$ and increases monotonically with momentum.
The total defect density $n$, Fig.~\ref{fig:fig4}b, exhibits power-law behaviour fitted to $\alpha=1.80(5)$, close the Kibble-Zurek prediction, $\alpha=5/3$, while error bars are larger compared to the lower-order EPs.
%

~\\
\noindent
{\bf Discussion}
\\
\noindent
Our work constitutes the first experimental investigation of non-Hermitian Kibble-Zurek scaling due to near-adiabatic passage across exceptional points.
The  programmability of our interferometric photon network allows us to engineer both Hermitian and non-Hermitian band structures with high tunability.
In particular, we demonstrated that we can also generate multi-band models which has enabled us  to access higher-order exceptional points and their distinctive properties.
As a key consequence, we have identified a platform for realizing non-Hermitian dynamics of increasing complexity as a route towards the many-body realm with
higher dimensional time evolution matrices.


\section{Methods}

\subsection{Theoretical background}

As detailed in Ref.~\cite{dhm}, we consider the time evolution of a (non-)Hermitian Hamiltonian $H(t)$.
The initial state is the ground state of the starting Hamiltonian, which is always Hermitian.
At time $t=0$ we start our time-dependent protocol, governed by
\begin{gather}
i \partial_t | \Psi(t) \rangle = H(t) | \Psi(t) \rangle,
\label{timedepsch}
\end{gather}
with $|\Psi(t)\rangle = \otimes_p |\Psi_p(t) \rangle$ for a given mode represented by momentum $p$, and the time dependence of $H(t)$ comes from $\Gamma(t)$ and $\Delta(t)$, with $\Gamma(t=0)=0$ always.

Under the non-unitary time evolution, the norm of the wave function is not conserved, so  an additional prescription for performing measurements in such states has to be given.
When interpreting such dynamics as a result of dissipation in the framework of a Lindblad master equation with an additional continuous measurement, expectation values of an operator $\mathcal{O}$ have to be evaluated as~\cite{ashida18,graefe2008,carmichael}
\begin{gather}
\langle \mathcal{O}(t)\rangle=\frac{\langle\Psi (t)|  \mathcal{O} |\Psi(t)\rangle}{\langle\Psi(t)|\Psi(t)\rangle} \, ,
\label{expvalue}
\end{gather}
where the left state, $\langle\Psi (t)|$ is taken as the Hermitian conjugate of the time evolved right state, $|\Psi(t)\rangle$.
Since the initial condition at $t=0$ is chosen to be the ground state of a Hermitian system, the initial right and left states also satisfy this condition.
In the following, we quantify the defect production via
\begin{equation}
        n = \frac{1}{\mathcal N} \sum_p \sigma_z(p), \,\,  \sigma_z(p) =\frac{\langle\Psi_p (t)|  \sigma_z  |\Psi_p(t)\rangle}{\langle\Psi_p(t)|\Psi_p(t)\rangle}
\end{equation}
with $\mathcal N$ denoting the number of considered momentum states.

The $\tau\rightarrow\infty$ limit of the momentum-resolved defect density coincides~\cite{dhm} with calculating the expectation value of $\sigma_z$
for a given $p$ momentum state
using the normalized right eigenfunction of the final Hamiltonian, corresponding to the smallest eigenvalue for the $\mathcal{PT}$-symmetric case,
or to the complex eigenvalue with the largest imaginary part otherwise. This defines $\sigma_z^{eq}(p)$. Upon integrating this over momenta, we obtain $n_{eq}$.

The density of states of Eq. \eqref{ham2x2} with $\Delta=\Gamma=0$ is defined as $\rho(E)=\frac{1}{\mathcal{N}}\sum_{p,\alpha=\pm}\delta(E-E_\alpha(p))$ with
$E_\pm(p)=\pm|p|$ and $\delta(E)$ the Dirac delta function. This becomes energy, $E$, independent and takes the value quoted in the main text.

\subsection{Experimental realisation of (non-)unitary time evolution for the two-level system}

In order to simulate the time evolution of these two-level systems, we encode the basis states in the horizontal and vertical polarisations of a single photon, with $\ket{H}=(1,0)^\text{T}$ and $\ket{V}=(0,1)^\text{T}$.
We generate heralded single photons via type-I spontaneous parametric down-conversion, with one photon serving as the trigger and the other as the signal.
The signal photon is initialized in the ground state $\ket{\Psi_p(0)}$ of $H_p$ with $\Gamma=0$ via a polarising beam splitter (PBS), a quarter-wave plate
(QWP) and a half-wave plate (HWP), with $p$-dependent parameters. We then send the single photon to the interferometric network as illustrated in Fig.~\ref{fig:setup}.

To simulate the non-unitary dynamics, driven by a time-dependent $H(t)$ up to a time $\tau$, we decompose the dynamics into
different momentum sectors, and directly implement the time-evolution operator $U_p(\tau)$. Specifically, we first numerically calculate $U_p(\tau)$ through
\begin{align}
U_p(\tau)=\prod_{k=1}^N e^{-iH_p(t_k)\delta t},\label{eq:UpT}
\end{align}
where $t_k=(k-1/2)\delta t$, $\delta t=\tau/N$, with $N\in \mathbb{N}$. We assume $H_p(t_k)$ to be time-independent within each $k$, and take sufficiently large $N$, such that Eq.~(\ref{eq:UpT}) converges.

As illustrated in Fig.~\ref{fig:setup}, we implement $U_p(\tau)$ according to
\begin{equation}
U_p=R(\theta_2,\varphi_2,\vartheta_2)L(\theta_H,\theta_V)R(\theta_1,\varphi_1,\vartheta_1),
\label{eq:U22}
\end{equation}
where the rotation operator $R(\theta_j,\varphi_j,\vartheta_j)$ ($j=1,2$) is realised using a sandwich-type wave-plate set, including a HWP at the setting angle $\varphi_j$, and two QWPs at $\theta_j$ and $\vartheta_j$,
respectively. The polarisation-dependent loss operator $L=\begin{pmatrix}0 & \sin2\phi_V\\ \sin2\phi_H&0 \end{pmatrix}$  is realised by a combination of two beam displacers (BDs) and two HWPs with setting angles $\phi_H$
and $\phi_V$. The setting angles $\{\theta_j,\varphi_j,\vartheta_j,\phi_H,\phi_V\}$ are fixed according to the numerically calculated $U_p(\tau)$. We note that Eq.~(\ref{eq:U22}) enables us to implement arbitrary
non-unitary operators for a two-level system with different setting angles~\cite{xiao}.

After performing the time evolution, we measure the expectation value of $\sigma_z$ through projective measurements.
Specifically, we measure the probability of photons in the
basis $\{\ket{H},\ket{V}\}$ through a PBS and avalanche photodiodes (APDs). The outputs are recorded in coincidence with trigger photons.
Typical measurements yield a maximum of $160,000$ photon counts. We then construct the momentum-resolved defect density through
\begin{equation}
\sigma_z(p,\tau)=\frac{\langle\Psi_p(\tau)|\sigma_z|\Psi_p(\tau)\rangle}{\langle\Psi_p(\tau)|\Psi_p(\tau)\rangle}=\frac{N_H-N_V}{N_H+N_V},
\end{equation}
where $N_H$ and $N_V$ are the photon counts with horizontal and vertical polarisations, respectively.

\subsection{Spectrum of $H^{(4)}_p$ with a fourth-order EP}

\begin{figure}
\includegraphics[width=0.5\textwidth]{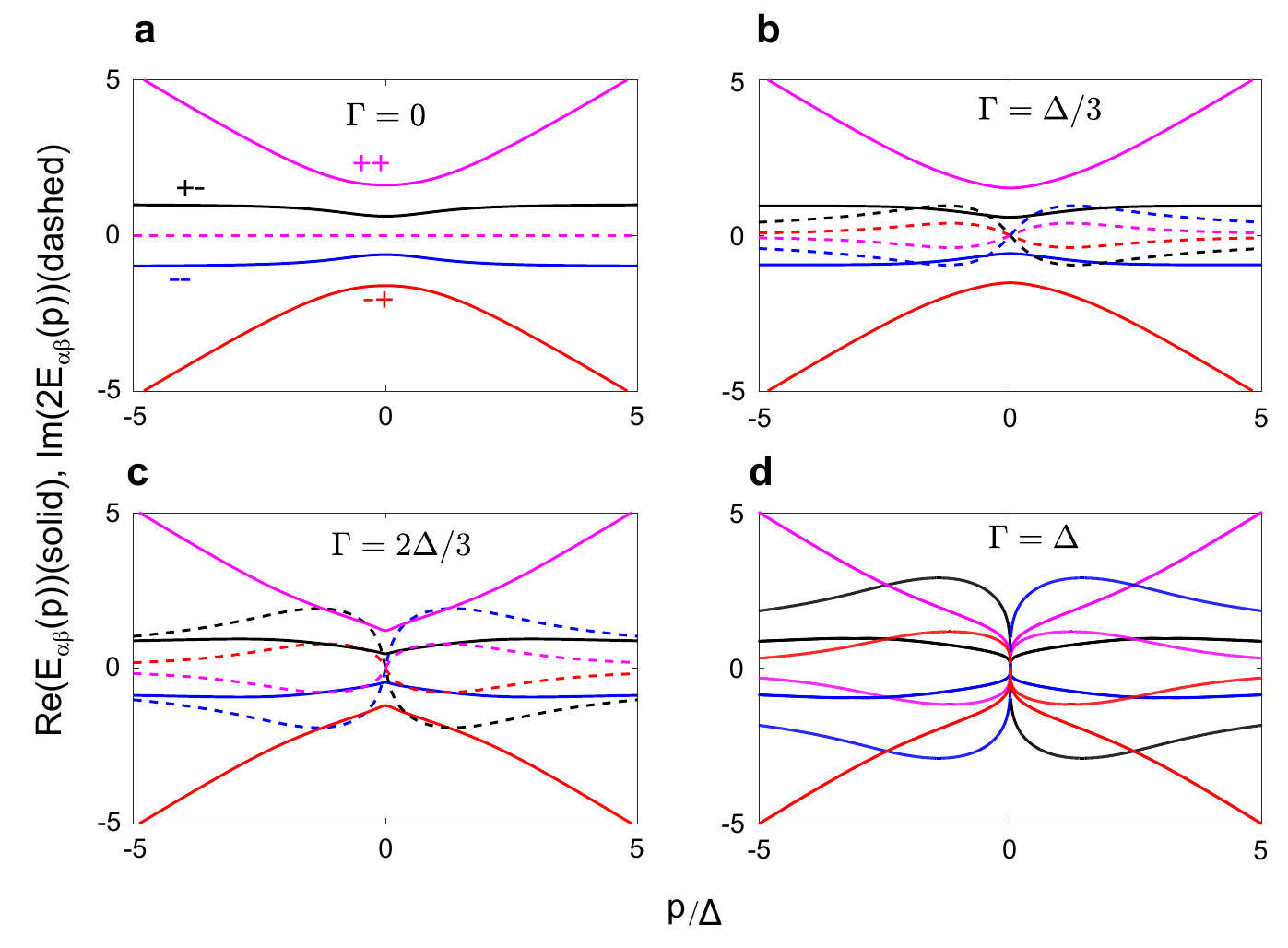}
\caption{{\bf The evolution of the instantaneous
spectrum of $H^{(4)}$.} The real and imaginary parts as a function of $\Gamma$ are shown in solid and dashed curves, respectively. The evolution starts from the Hermitian case with $\Gamma=0$ {\bf a}, through the $\mathcal{PT}$-broken regions {\bf b} and {\bf c}, and ends at $\Gamma=\Delta$ {\bf d}, where the EP4 appears are at $p=0$.
}
\label{fig:specevolep4}
\end{figure}

The 4x4 Hamiltonian, $H^{(4)}_p$ in Eq. \eqref{ep4ham} can be diagonalized analytically, yielding four bands
as
\begin{widetext}
\begin{gather}
E_{\alpha\beta}(p)=\alpha \frac 12 \sqrt{2\beta \sqrt{p^4-2\Gamma^2p^2+2p^2\Delta^2+5\Gamma^4-10\Gamma^2\Delta^2+5\Delta^4+8ip\Gamma^3+24i\Gamma p\Delta^2}+2p^2-6\Gamma^2+6\Delta^2},
 \end{gather}
\end{widetext}
with $\alpha=\pm$, $\beta=\pm$. The evolution of the instantaneous spectrum is depicted in Fig.~\ref{fig:specevolep4}. For $\Gamma=\Delta$, it reduces to
\begin{gather}
E_{\alpha\beta}(p)=\frac{\alpha}{\sqrt 2}\sqrt{p^2+\beta\sqrt{p^4+32i\Delta^3p}},
\end{gather}
which is dominated by the term $32i\Delta^3p$ under the double square root for small momenta, thus realising an EP4.

As advertised above, this Hamiltonian can also be rewritten in terms of the Pauli matrices
of two interacting spins, denoted by $\tau_i$ and $\sigma_i$ with $i=x$, $y$, $z$.
Such a minimal many-body model reads
\begin{gather}
H^{(4)}_p=ip(\tau_+\sigma_+ -\tau_-\sigma_-) +\Delta\sigma_x-i\Gamma\sigma_y+\nonumber\\
+\tau_+\sigma_-(\Delta-\Gamma)+\tau_-\sigma_+(\Delta+\Gamma).
\end{gather}

\subsection{Time evolution of the four-level system.}

Similar to the case with two-level systems, we numerically calculate the time-evolution operator
$U^{(4)}_p(\tau)$ at the final time $\tau$ for each $p$-sector. More explicitly, we calculate $U^{(4)}_p(\tau)$ through $U^{(4)}_p=\prod_{k=1}^N e^{-iH^{(4)}_p(t_k)\delta t}$, for sufficiently large $N$.

To simulate the dynamics of the four-level system governed by $H^{(4)}$, we encode the four basis states in the polarisations and spatial modes of single photons, with the basis states given
by $\{|UH\rangle=(1,0,0,0)^\text{T},|UV\rangle=(0,1,0,0)^\text{T},|DH\rangle=(0,0,1,0)^\text{T},|DV\rangle=(0,0,0,1)^\text{T}\}$. Here $|U\rangle$ and $|D\rangle$ represent, respectively,
the upper and lower spatial modes of photons (see Fig.~\ref{fig:setup}b).
The experimental implementation of the calculated $U^{(4)}_p$, however, is different from that of $U_p$ for two-level systems which is realised {\it exactly} by BDs and wave plates. Instead,
we approximate $U^{(4)}_p$ with a series of modules, each consisting of two BDs and a set of wave plates. Specifically, each module features two sets of sandwiched
wave plates (QWP-HWP-QWP), and a sandwiched set of BDs and wave plates (BD-HWP-BD).
The QWP-HWP-QWP configuration realises a controlled-rotation $|U\rangle\langle U|\otimes R_U+|D\rangle\langle D|\otimes R_D$, and
the BD-HWP-BD structure introduces the non-unitary operator
\begin{align}
L^{(4)}=\begin{pmatrix}
             0 & \sin2\phi_U & 0 & 0 \\
             \sin2\phi_M & 0 & 0 & -\cos2\phi_M \\
             \cos2\phi_M & 0 & 0 & \sin2\phi_M \\
             0 & 0 & \sin2\phi_L & 0
           \end{pmatrix},
\end{align}
where $\phi_M$ is the setting angle of the HWPs between two BDs.

\begin{figure}
\includegraphics[width=0.4\textwidth]{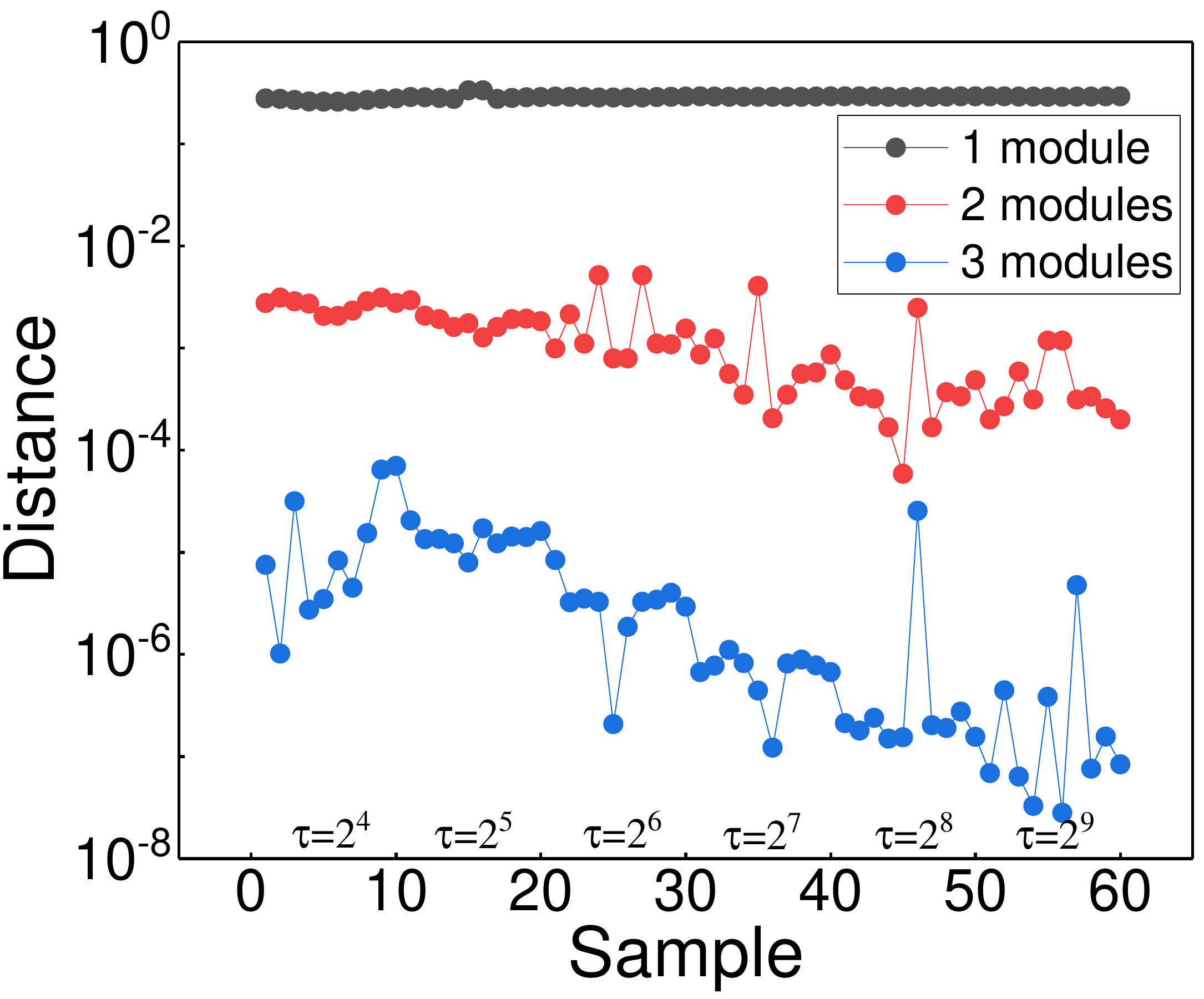}
\caption{{\bf Distance between the implemented non-unitary operators and the ideal ones.} Calculated distance $d$ for all sampled $p$ and $\tau$, with different number of modules in the implementation of $U^{(4)}_p$. We plot the distance for all sampled $p$ with $\tau\Delta=2^n$ in the $x$-axis range $(10n-40,10n-30)$.
}
\label{fig:dist}
\end{figure}

To estimate the deviation of the implemented non-unitary time-evolution operator $U$ with respect to the ideal $U^{(4)}_p$, we define the distance
\begin{equation}
d:=1-\frac{\left|\text{Tr}\left[\tilde{U}\tilde{U}^{(4)\dagger}_p\right]\right|}{\sqrt{\text{Tr}\left[\tilde{U}\tilde{U}^\dagger\right]\text{Tr}\left[\tilde{U}^{(4)}_p\tilde{U}^{(4)\dagger}_p\right]}},
\end{equation}
where $\tilde{U}:=U/\text{Tr}\left[UU^\dagger\right]$. The distance varies between $0$ and $1$, with $0$ indicating a perfect implementation of $U^{(4)}_p$. As shown in Fig.~\ref{fig:dist}, the distance
is already below $10^{-4}$ for the experimentally relevant parameters when we use three sets of modules. In principle, we can achieve even smaller distances by increasing the number of modules.

In order to test the Kibble-Zurek scaling around EP4 experimentally,
we probe $M_y(p)$ via projective measurements of the time-evolve state $|\Psi_p(\tau)\rangle$. More specifically,
we implement the transformation $O=|UH\rangle\langle o_1|+|UV\rangle\langle o_2|+|DH\rangle\langle o_3|+|DV\rangle\langle o_4|$ through
cascaded interferometers consisting of BDs and wave plates. Here $|o_i\rangle$ ($i=1,\cdots, 4$) are eigenstates of $M_y$, with
$M_y=\sum_i o_i|o_i\rangle\langle o_i|$, where $o_i$ are the eigenvalues of $M_y$. A PBS is then used to map the basis states $\{|UH\rangle,|UV\rangle, |DH\rangle,|DV\rangle\}$ to four distinct spatial modes,
where photons are collected by four APDs. It follows that $M_y(p)=\sum_{i=1}^{4} o_i P_i$, where $P_i$ is the photon count from the corresponding APD.
We note that $M_y^{\text{eq}}$ is measured in a similar fashion, by performing projective measurements directly on the state $|\varphi_p\rangle$.

\subsection{Error analysis.}

The difference between the experimental data and theoretical predictions is mostly caused by the inaccuracy of the wave-plate parameters, as well as the dephasing of BDs in the cascaded interferometric network. Since both wave plates and BDs are tuned independently in different $p$-sectors, experimental errors in different $p$-sectors are uncorrelated.

For the data shown in Fig.~\ref{fig:fig2}a, the unitary time evolution is realised by three HWPs and without any interferometers (hence no BDs). In contrast, the data shown in Figs.~\ref{fig:fig2}b, \ref{fig:fig3}a and \ref{fig:fig4}a, the non-unitary time evolutions therein require more wave plates as well as BDs. This is the reason for the more apparent experimental imperfections in the case of non-unitary time evolutions.
However, since the imperfections in each $p$ are independent (see discussion above), and the data shown in Figs.~\ref{fig:fig2}c, \ref{fig:fig3}b and \ref{fig:fig4}b are obtained by integrating over $p$, the differences between the experimental data and theoretical predictions are smaller for the total defect densities.

{\bf Acknowledgments}
This research is supported by the Natural Science Foundation of China (Grant Nos.\ 11674056, 11674189, U1930402, 11974331) and the startup fund from Beijing Computational Science Research Centre. We also acknowledge support from
the National Research, Development and Innovation Office - NKFIH   within the Quantum Technology National Excellence Program (Project No.
      2017-1.2.1-NKP-2017-00001), K119442 and by
 a grant from the Simons Foundation.
This work was performed in part at Aspen Center for Physics, which is supported by National Science Foundation grant PHY-1607611.
Support by the Deutsche Forschungsgemeinschaft via the  Leibniz  Prize  program, SFB 1143 and cluster of excellence EXC2147 ct.qmat (project-ids 247310070, 39085490)
is also acknowledged.
This project has received funding from the European Research Council (ERC) under the European Union's Horizon 2020 research and innovation programme (grant agreement No. 853443).
WY acknowledges support from the National Key Research and Development Program of China (Grant Nos. 2016YFA0301700 and 2017YFA0304100).

{\bf Author contributions}
PX designed the experiments, and analysed the results with contributions from HL, JD and WY. LX and DQ performed the experiments with contributions from KW. BD, MH and RM developed the theoretical aspects and performed the theoretical analysis. PX, BD, MH, RM, and WY supervised the project and wrote the manuscript with input from other authors.

{\bf Additional information}
Correspondence and requests for materials should be addressed to Peng Xue (gnep.eux@gmail.com).

{\bf Competing financial interests}
The authors declare no competing financial interests.

{\bf Data availability}
The data represented in Figs. 2-4 and other findings of this study are available from the corresponding author upon reasonable request.

\end{document}